\documentclass[reprint,prd]{revtex4-1}

\usepackage{graphicx}
\usepackage[T2A]{fontenc}
\usepackage[utf8]{inputenc}

\usepackage{amsmath}
\usepackage{amsfonts}
\usepackage{makeidx}
\usepackage{mathrsfs}

\usepackage{titleps}

\begin{document}

\pagenumbering{gobble}

\title{Moving unstable particles and special relativity }

\author{Eugene V. Stefanovich}
\email{$eugene\_stefanovich@usa.net$}
\affiliation{Mountain View, California, USA}

\begin{abstract}
In Poincar\'e-Wigner-Dirac theory of relativistic interactions, boosts are dynamical.  This means that -- just like time translations -- boost transformations have non-trivial effect on internal variables of interacting systems.   This is different from space translations and rotations, whose actions are always universal, trivial and interaction-independent.  Applying this theory to unstable particles viewed from a moving reference frame, we prove that the decay probability cannot be invariant with respect to boosts. Different moving observers may see different internal compositions of the same unstable particle.  Unfortunately, this effect is too small to be noticeable in modern experiments.
\end{abstract}


\maketitle

\section{Introduction}

Time dilation is one of the most spectacular predictions of special relativity.  It means that any time-dependent process  slows down by the universal factor of $ 1/\sqrt{1 - v^2/c^2} \equiv \cosh \theta$, when viewed from a reference frame moving with the speed $v$ \footnote{In relativistic formulas, it is often convenient to use the rapidity parameter $\theta$ defined so that $v = c \tanh \theta$, where $c$ is the speed of light, and $ 1/\sqrt{1 -v^2/c^2} = \cosh \theta \geq 1$.}.  The textbook example of such a time-dependent process is the decay law $\Upsilon (0,t)$ of an unstable particle at rest. The function $\Upsilon (0,t)$ is the probability of finding the unstable particle at time $t$, if it was prepared with 100\% certainty at time $t=0$. Then, according to special relativity, the decay law of a moving particle should be $\cosh \theta$ times slower

\begin{align}
\Upsilon^{SR}(\theta, t) &= \Upsilon (0,t/\cosh \theta)  \label{eq:ups-theta}
\end{align}

\noindent Indeed, this prediction
was confirmed in numerous measurements \cite{dilation, Frisch, dilation1, dilation2}. The best accuracy of 0.1\% was achieved in experiments with relativistic muons \cite{muons, muons2}.

However, the exact validity of (\ref{eq:ups-theta}) is still a subject of controversy.  One point of view \cite{Khalfin, Urbanowski3, Fleming} is that special-relativistic time dilation was derived in the framework of classical theory and may not be directly applicable to unstable particles, which are fundamentally quantum systems without well-defined masses, velocities, positions, etc.

\begin{quote}
\emph{However,  such  a  quantum  clock  as  an  unstable  particle  cannot  be  at  rest
(i.e., cannot have zero velocity or zero momentum) and simultaneously be at a definite point (due to the quantum uncertainty relation). So, the standard derivation
of the moving clock dilation is inapplicable for the quantum clock. The related
quantum-mechanical derivation must contain some reservations and corrections.} M. I. Shirokov \cite{Shirokov_decay}
\end{quote}

Indeed, detailed quantum-mechanical calculations \cite{Stefanovich_decay, Stefanovich_dec, Shirokov_decay, Urbanowski3} suggest that (\ref{eq:ups-theta}) is not accurate, and that corrections to this formula should be expected, especially at large times  exceeding multiple lifetimes ($t \gg 1/\Gamma$). Although, these corrections are too small to be observed in modern experiments, their presence casts doubt on the limits of applicability of Einstein's special relativity.

Unfortunately, the results \cite{Stefanovich_decay, Stefanovich_dec, Shirokov_decay, Urbanowski3} were derived under certain assumptions and approximations. So, the question remains whether one can design a relativistic model in which the decay slowdown will acquire exactly the form (\ref{eq:ups-theta}) demanded by special relativity \cite{Exner, Alavi}?

In order to answer this question we will analyze the status of interactions in special relativity from a more general point of view. We are going to prove that under no circumstances the decay law of a moving particle transforms under boosts exactly as in  (\ref{eq:ups-theta}).

\section{Materials and Methods}
\label{sc:special}

\subsection{Inertial transformations}
\label{ss:inertial}

The theory of relativity is supposed to connect views of different inertial observers.  The principle of relativity says that all such observers are equivalent, i.e., two inertial observers performing the same experiment will obtain the same results.

There are four classes of inertial transformations -- space translations, time translations, rotations and boosts -- and their actions on observed systems differ very much (see Table \ref{table:7.1}).  For example, describing results of space translations and rotations is very easy. An observer displaced by the 3-vector $\boldsymbol{a}$ sees all atoms in the Universe simply shifted in the opposite direction $-\boldsymbol{a}$.  This shift is absolutely exact and universal.  It applies to all systems, however complicated. The same can be said about rotations.  One can switch to the point of view of the rotated observer by simply rotating all atoms in the Universe. For example, rotation through the angle $\varphi$ about the $z$-axis  results in the following transformation of coordinates

\begin{widetext}
\begin{table}[h]
\caption{Inertial transformations.}
\begin{tabular*}{\textwidth}{@{\extracolsep{\fill}}rcccl}
\hline
Transformation            & Type & Parameter & Generator &  Meaning of generator \cr
\hline
Space translation   & kinematical & distance $\boldsymbol{a}$ & $\boldsymbol{P} = \boldsymbol{P}_0$    &  Total momentum \cr
Rotation   & kinematical  & angle $\boldsymbol\varphi$  & $\boldsymbol{J} = \boldsymbol{J}_0$ &  Total angular momentum \cr
Time translation   &  dynamical & time $t$ & $H= H_0+V$  &  Total energy (Hamiltonian) \cr
Boost   & dynamical & rapidity $\boldsymbol \theta$ & $\boldsymbol{K}= \boldsymbol{K}_0 + \boldsymbol{Z}$   & Boost operator \cr
\hline
\end{tabular*}
\label{table:7.1}
\end{table}
\end{widetext}

\begin{align}
x' &= x \cos \varphi - y \sin \varphi \label{eq:rotx} \\
y' &= y \cos \varphi + x \sin \varphi  \\
z' &= z  \label{eq:rotz}
\end{align}

\noindent  which is independent on the composition of the observed system and on its interactions.
Due to this exact universality, we can regard space translations and rotations as purely geometrical or \emph{kinematical} transformations.

Time translation is also an inertial transformation, because repeating the same experiment at different times will not change the outcome.  However, this transformation is by no means kinematical. Time evolutions of interacting systems are very complicated.  Their descriptions require intimate knowledge of the systems' composition, state and interactions acting between systems' parts.  We will say that time translations are \emph{dynamical} transformations.

Now, what about boosts? Are they kinematical or dynamical? In non-relativistic classical physics boosts are definitely regarded as kinematical -- they simply change velocities of all atoms in the Universe.  However, things become more complicated in relativistic physics, as we will see below.

\subsection{Boosts in special relativity}
\label{ss:boost-special}

Description of boost transformations is the central subject of special relativity. Einstein based his approach on the already mentioned relativity postulate and on his second postulate about the invariance of  the speed of light.  It is remarkable how all results of special relativity can be derived from these two simple and undeniable statements.

\begin{figure}
\includegraphics[width=7cm,height=2.6cm] {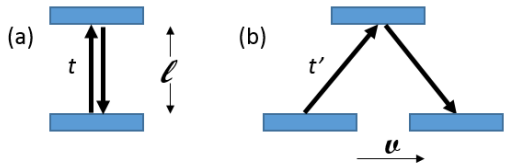} \caption{Light clock: (a) at rest, (b) in motion perpendicular to the clock's axis.}
\label{fig:light-clock1}
\end{figure}

Consider the light clock shown in Fig. \ref{fig:light-clock1}(a).  It consists of two parallel mirrors and the light pulse reflecting back an forth between them. The period of the clock at rest is equal to $\tau = 2t = 2l/c$.  If the clock is moving, as in Fig. \ref{fig:light-clock1}(b), the distance traveled by the light pulse increases to $l' =2ct' =  2\sqrt{l^2 + \left(vt'\right)^2}$.  Solving this system of equations with respect to the clock period we obtain

\begin{align}
\tau' = 2t' = \tau/\sqrt{1 -v^2/c^2} = \tau  \cosh \theta \label{eq:period}
\end{align}

\noindent So, the moving clock goes $\cosh \theta$ times slower than the clock at rest. This is the time dilation effect that we used in Eq. (\ref{eq:ups-theta}).

\begin{figure}
\includegraphics[width=7cm,height=5cm] {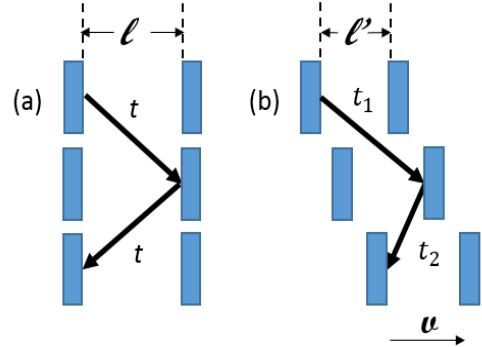} \caption{Light clock: (a) at rest and (b) in motion parallel to the clock's axis. The time evolution is shown in three frames stacked up vertically.}
\label{fig:light-clock2}
\end{figure}

Let us now consider the same clock oriented parallel to its velocity, as in Fig. \ref{fig:light-clock2}.  The clock's rate should not depend on its orientation, so we already know the period of the moving clock (\ref{eq:period}).  Assuming the invariance of the speed of light, this period can be achieved only if the distance between the two mirrors decreases.  The corresponding system of equations is

\begin{align*}
\tau' = t_1 + t_2 = (l' + vt_1)/c + (l' - vt_2)/c
\end{align*}

\noindent Solving with respect to $l'$, we obtain the familiar length contraction formula

\begin{align}
l' = l\sqrt{1 -v^2/c^2} = l /\cosh \theta   \label{eq:length}
\end{align}

Formulas (\ref{eq:period}) and (\ref{eq:length}) already imply that no material object can move faster than the speed of light. Otherwise, the factor $\sqrt{1 -v^2/c^2}$ would become imaginary, which is absurd.

We can also make a clock, in which, instead of the light pulse, we have  a massive steel ball reflecting between the two mirrors.  The ball's speed $w$ is less than $c$, and the speed invariance postulate does not apply to $w$.  Nevertheless, we expect this clock to obey the same time dilation and length contraction rules as derived above.  Then, for consistency, we have to modify the velocity transformation law.  For example, if the resting clock in Fig. \ref{fig:light-clock2}(a)  had ball's velocities $\pm w$, then the moving clock in Fig. \ref{fig:light-clock2}(b)  should have velocities

\begin{align*}
w_1 &= \frac{w + v}{1 + wv/c^2} \\
w_2 &= \frac{-w + v}{1 - wv/c^2}
\end{align*}

\noindent  Indeed, it is not difficult to verify that these values solve the system of equations

\begin{align*}
w_1 t_1 &= l' + vt_1 \\
-w_2 t_2 &= l' - vt_2 \\
t_1 + t_2 &= (2l /w) \cosh \theta
\end{align*}

\noindent that describe the movement of the ball during one clock period.

Special relativity textbooks explain how all these particular results follow from Lorentz transformation formulas for the times and positions of events.  For example, any event having 4-coordinates $(t,x,y,z)$ in the rest frame, will have other 4-coordinates

\begin{align}
 t' &=  t \cosh \theta - (x/t) \sinh \theta \label{eq:lorentz-transform-t} \\
 x' &= x \cosh \theta - ct \sinh \theta \label{eq:lorentz-transform-x} \\
 y' &= y \\
 z' &= z \label{eq:lorentz-transform-z}
 \end{align}

\noindent in the frame moving with velocity $v = c \tanh \theta$ along the $x$-axis.  The linear character and the exact universality of these formulas reminds transformations of 3-coordinates under rotations (\ref{eq:rotx}) -- (\ref{eq:rotz}). So, it is tempting to continue this analogy and to introduce the idea of the 4D Minkowski space-time, whose points constitute physical events, and where boosts are represented by purely geometrical (kinematical) pseudo-rotations.

However, it is important to note that all the above derivations used model systems without interactions. Of course, reflections of light pulses or steel balls from the mirrors do involve interactions, but in our idealized thought experiments we can assume that these processes take negligibly short times. The second Einstein's postulate is, actually, applicable only to light pulses and events associated with them. So, strictly speaking, we are not allowed to extend results of special relativity beyond corpuscular optics.  However, one can show \cite{Stefanovich_Mink} that Lorentz transformations (\ref{eq:lorentz-transform-t}) -- (\ref{eq:lorentz-transform-z}) can be also extended to events -- such as intersections of particle trajectories -- involving massive non-interacting particles, e.g., our steel ball and mirrors.

How can we be confident that the same conclusions apply to interacting systems?  For example, what if the steel ball is bouncing between plates of a charged capacitor?  Can we be sure that Lorentz formulas (\ref{eq:lorentz-transform-t}) -- (\ref{eq:lorentz-transform-z}) still apply?

Here we meet the following fork in the road. On one hand, we can postulate that the laws of special relativity established above are valid independent of interactions. Then boosts should be rigorously kinematical, just as space translations and rotations.  This non-obvious postulate is tacitly assumed in all textbooks. In particular, it was used in numerous attempts \cite{Lee-Kalotas, Levy-Leblond, Sardelis, Schwartz, Field, Polishchuk, Galiautdinov} to derive Lorentz transformations from the first Einstein postulate only.

Alternatively, we can assume that, similar to time translations, boosts are dynamical, i.e., they involve interactions, and their actions cannot be expressed by simple universal formulas, like (\ref{eq:lorentz-transform-t}) -- (\ref{eq:lorentz-transform-z}). We will discuss these two possibilities in section \ref{sc:dynamical}. However, before doing that, in the next three subsections, we are going to recall the fundamentals of relativistic quantum theory pioneered by Wigner and Dirac. This theory is based on the extremely important fact that inertial transformations form the Poincar\'e group.

\subsection{Representations of the Poincar\'e group in quantum mechanics}

We are interested in application of inertial transformations to quantum systems. Properties of such systems are described by objects in the Hilbert space $\mathscr{H}$, e.g., state vectors (wave function) and Hermitian operators of observables.  So, we have to define the action (or representation) of inertial transformations in $\mathscr{H}$. Operators of this representation $U_g$ must preserve quantum-mechanical probabilities, so these operators have to be unitary \cite{Wigner-eng}.  This brings us to the classical mathematical problem of constructing unitary representations $U_g$ of the Poincar\'e group in the given Hilbert space $\mathscr{H}$  \cite{Wigner_unit}.

Important role is played by the so-called ``infinitesimal transformations'' or generators.  They are represented by  Hermitian operators in $\mathscr{H}$ (see Table \ref{table:7.1}). Unitary representatives $U_g$ of finite transformations can be expressed by exponential functions of the Hermitian generators.  For example, a general inertial transformation consisting of (boost $\boldsymbol\theta$) $\times$ (rotation $\boldsymbol\varphi$) $\times$ (space translation    $\boldsymbol{a}$) $\times$ (time translation $t$) is represented by the following product of unitary exponents

\begin{align*}
U_g = e^{-\frac{ic}{\hbar}\boldsymbol{K} \cdot \boldsymbol\theta}e^{-\frac{i}{\hbar}\boldsymbol{J} \cdot\boldsymbol\varphi}
e^{-\frac{i}{\hbar}\boldsymbol{P} \cdot \boldsymbol{a}} e^{\frac{i}{\hbar}Ht}
\end{align*}

Commutators of the Hermitian generators are fully determined by the structure of the Poincar\'e group \cite{Dirac, book}

\begin{align}
\left[J_i, P_j\right] &= i\hbar \sum_{k=1}^3 \epsilon_{ijk} P_k
\label{eq:5.50} \\
\mbox{ } \left[J_i, J_j\right] &= i\hbar \sum_{k=1}^3 \epsilon_{ijk} J_k
\label{eq:5.51} \\
\mbox{ } \left[J_i, K_j\right] &= i\hbar \sum_{k=1}^3 \epsilon_{ijk} K_k
\label{eq:5.52} \\
\mbox{ } \left[P_i,P_j\right] &=  \left[J_i,H\right] = \left[P_i, H\right] = 0 \label{eq:5.53} \\
 \mbox{ } \left[K_i, K_j\right] &= -\frac{i\hbar}{c^2} \sum_{k=1}^3 \epsilon_{ijk} J_k
\label{eq:5.54} \\
\mbox{ } \left[K_i, P_j\right] &= -\frac{i\hbar}{c^2} H \delta_{ij} \label{eq:5.55} \\
 \mbox{ } \left[K_i, H\right] &= -i\hbar P_i
\label{eq:5.56}
\end{align}

\subsection{Hilbert space of unstable particle}

According to Wigner \cite{Wigner_unit, book}, the Hilbert space $\mathscr{H}^{(i)}$ of each stable elementary particle carries an unitary irreducible representation $U_g^{(i)}$ of the Poincar\'e group.  The Hilbert space of an $N$-particle system is constructed as a tensor product (with proper (anti)symmetrization) of one-particle spaces

\begin{align}
\mathscr{H}^{N} = \mathscr{H}^{(1)} \otimes \mathscr{H}^{(2)} \otimes \ldots \otimes \mathscr{H}^{(N)}  \label{eq:HN}
\end{align}

\noindent  In the formalism with varied numbers of particles (e.g., in quantum field theory), one builds the Fock space as the direct sum of spaces (\ref{eq:HN}) with fixed numbers of particles.  For example, in a good approximation one can describe the unstable particle $\alpha$ with one decay channel $\alpha \to \beta + \gamma$, in the part of the Fock space, which includes the particle $\alpha$ itself and its decay products $\beta + \gamma$

\begin{align*}
\mathscr{H} = \mathscr{H}^{(\alpha)} \oplus \left(\mathscr{H}^{(\beta)} \otimes \mathscr{H}^{(\gamma)} \right)
\end{align*}

The probability of finding the unstable particle $\alpha$ in any state is defined as the square of the projection of the state vector $| \Psi \rangle \in \mathscr{H}$ on the subspace $\mathscr{H}^{(\alpha)}$.  Equivalently, this probability can be expressed as the expectation value of the Hermitian projection $T$ on the subspace $\mathscr{H}^{(\alpha)}$

\begin{align*}
T \mathscr{H}^{(\alpha)} &= \mathscr{H}^{(\alpha)} \nonumber \\
\Upsilon &= \langle \Psi | T | \Psi \rangle
\end{align*}

\noindent  Then, the decay law --  the time evolution of the probability $\Upsilon$ -- is obtained as

\begin{align}
\Upsilon (0,t) = \left\langle \Psi \left| e^{\frac{i}{\hbar}Ht} T e^{-\frac{i}{\hbar}Ht} \right| \Psi \right\rangle \label{eq:decay_law}
\end{align}

\subsection{Interacting representation of the Poincar\'e group}

To perform decay calculations with formula (\ref{eq:decay_law}), we have to specify the Hamiltonian $H$ in the Hilbert space $\mathscr{H}$.  In order to keep the relativistic invariance, this Hamiltonian should be consistent with other Poincar\'e generators, i.e., commutation relations (\ref{eq:5.50}) -- (\ref{eq:5.56}) have to be satisfied.

Using available 1-particle irreducible representations $U_g^{(\alpha)}, U_g^{(\beta)}, U_g^{(\gamma)}$, one can easily construct one valid representation of the Poincar\'e group in   $\mathscr{H}$

\begin{align*}
U_g^0 \equiv U_g^{(\alpha)} \oplus \left(U_g^{(\beta)} \otimes U_g^{(\gamma)} \right)
\end{align*}

\noindent It is appropriate to call this representation ``non-interacting,''  because its generators $\{\boldsymbol{P}_0, \boldsymbol{J}_0, \boldsymbol{K}_0, H_0  \}$ take the form corresponding to free particles. Apparently, in this case, the subspace $ \mathscr {H} _{\alpha} $ remains invariant with respect to all inertial transformations. In particular,  non-interacting translation generators  commute with the projection $T$

\begin{align}
[T, H_0] &= 0 \nonumber \\
[T, \boldsymbol{P}_0] &= 0 \label{eq:commp0}
\end{align}

According to Dirac \cite{Dirac, book}, one can introduce relativistic interaction by defining  in $ \mathscr {H} $ a new unitary representation $U_g \neq U_g^0$ of the Poincar\'e group with generators $\{\boldsymbol{P}, \boldsymbol{J}, \boldsymbol{K}, H  \}$.  Referring to our understanding of the kinematical/dynamical character of transformations from subsection \ref{ss:inertial}, we can immediately conclude that generators of space translations and rotations coincide with their non-interacting counterparts

\begin{align}
\boldsymbol{P}= \boldsymbol{P}_0 \label{eq:kinematP} \\
\boldsymbol{J} = \boldsymbol{J}_0 \label{eq:kinematJ}
\end{align}

\noindent while the generator of time translations contains a  non-trivial interaction term $V$

\begin{align*}
H= H_0 +V
\end{align*}

\noindent (see Table \ref{table:7.1}). It is important to note that the Hermitian projection $T$ cannot commute with this interaction and with the total Hamiltonian $H$

\begin{align}
[T, H] = [T,V]  \neq 0 \label{eq:comm}
\end{align}

\noindent Indeed, only in this case, the decay law  is a non-trivial function of time

\begin{align*}
\Upsilon (0, 0) &= \langle \Psi | T | \Psi \rangle = 1 \\
\Upsilon(0,t>0) &= \left\langle \Psi \left| e^{\frac{i}{\hbar}Ht} T e^{-\frac{i}{\hbar}Ht} \right| \Psi \right\rangle < 1
\end{align*}

\noindent as required for any unstable particle.

A Poincar\'e-Wigner-Dirac relativistic quantum description of any isolated interacting system is constructed in a similar manner.  In the Hilbert space $ \mathscr {H} $ of the system one defines 10 Hermitian generators $\{\boldsymbol{P}_0, \boldsymbol{J}_0, \boldsymbol{K}_0 + \boldsymbol{Z}, H_0+ V  \}$ with commutators (\ref{eq:5.50}) -- (\ref{eq:5.56}).  These operators not only specify the basic total observables of the system, but also determine how the results of observations transform from one inertial system to another.  Moreover, one can switch to the classical relativistic description by taking the limit $\hbar \to 0$ and considering only states describable by localized quasiclassical wave packets, which can be approximated by points in the phase space.  In this limit, observables are replaced by real functions on the phase space, quantum commutators are represented by Poisson brackets,  and time evolution is approximated by trajectories in the phase space \cite{mybook, volume1}.

\section{Results and discussion}
\label{sc:dynamical}

\subsection{Kinematical boosts}

As we mentioned at the end of subsection \ref{ss:boost-special}, Einstein's special relativity assumes that boost transformations can be represented by exact Lorentz formulas (\ref{eq:lorentz-transform-t}) -- (\ref{eq:lorentz-transform-z}), which are valid universally for all events and physical systems, independent on their state, composition, and involved interactions. In other words, in special relativity boosts are kinematical.

In classical relativistic physics, this hypothesis is known as the  condition of ``invariant trajectories'' or ``manifest covariance''. The well-known Currie-Jordan-Sudarshan theorem \cite{CJS} states that this condition is not compatible with the  Hamiltonian description of dynamics presented in the previous section.  In other words, a Poincar\'e-invariant theory with invariant trajectories can exist only in the absence of interactions.
 This explains  the name ``no-interaction theorem'' often used for the Currie-Jordan-Sudarshan result.
Several options were tried in the literature for explaining this paradox.

One idea was that Hamiltonian dynamics is not suitable for describing relativistic interactions. Instead, various non-Hamiltonian theories were offered \cite{Van_Dam, Van_Dam2, Sudarshan83, covariant, Keister, Sokolov_mechanics}, which deviated  from  the  Poincar\'e-invariant Wigner-Dirac approach. So far, the predictive power of these theories remains rather limited.

Another idea is to abandon  particles and  replace them by (quantum) fields \cite{Malamentprd, Wilczek, Halvorson, Strocchi, Boyer-08}, because \emph{``there are no particles, there are only fields''} \cite{Hobson}.  This approach goes as far as claiming that there is no point in discussing such things as  observables (positions and momenta) of interacting particles, their wave functions, and also their time evolutions in the interacting regime.

\begin{quote}
\emph{The more one thinks about this situation, the more one is led to the conclusion that one should not insist on a detailed description of the system in time. From the physical point of view, this is not so surprising, because in contrast to non-relativistic quantum mechanics, the time behavior of a relativistic system with creation and annihilation of particles is unobservable. Essentially only scattering experiments are possible, therefore we retreat to scattering theory. One learns modesty in field theory.} G. Scharf \cite{Scharf}
\end{quote}

\begin{quote}
\emph{The foregoing discussion suggests that the theory will not consider the time dependence of particle interaction processes.  It will show that in these processes there are no characteristics precisely definable (even within the usual limitations of quantum mechanics); the description of such a process as occurring in the course of time is therefore just as unreal as the classical paths are in non-relativistic quantum mechanics.  The only observable quantities are the properties (momenta, polarizations) of free particles: the initial particles which come into interaction, and the final particles which result from the process (L. D. Landau and R. E. Peierls, 1930).} V. B. Berestetski{\u{\i}}, E. M. Livshitz and
L. P. Pitaevski{\u{\i}} \cite{BLP}
\end{quote}

We cannot accept this point of view, because it has nothing to say about such interacting time-dependent system as the unstable particle.

\subsection{Dynamical boosts}

Our preferred way to resolve the Currie-Jordan-Sudarshan controversy is to accept that  boost transformations are dynamical.
Actually, it was mentioned even in the original Dirac's paper \cite{Dirac} that in a theory with kinematical space translations (\ref{eq:kinematP}) and rotations (\ref{eq:kinematJ}), boosts must depend on interactions. Indeed, if we assume that boosts are kinematical ($\boldsymbol{Z} = 0$), then we obtain from (\ref{eq:5.55})

\begin{align*}
H &= \frac{i c^2}{\hbar} \left[K_x, P_x\right] = \frac{i c^2}{\hbar} \left[(K_0)_x, (P_0)_x\right] = H_0
\end{align*}

\noindent the absurd proposition that interaction in the Hamiltonian must vanish ($V \equiv H - H_0 = 0$).  Therefore, we should have $V \neq 0$, $\boldsymbol{Z} \neq 0$, which  means that we are working in the \emph{instant form} of dynamics, according to Dirac's classification \cite{Dirac}.

\subsection{Decays caused by boosts} \label{sc:dec-boost}

Our conclusion about the dynamical character of boosts disagrees with the usual special-relativistic ``geometrical'' view on boosts.  In particular, we can no longer claim that

\begin{quote}
\emph{Any event that is ``seen'' in one inertial system is ``seen''
in all others. For example if observer in one system ``sees'' an
explosion on a rocket then so do all other observers.} R. Polishchuk
\cite{Polishchuk}
\end{quote}

Returning to our example of unstable particle, we can say that when the observer at rest sees the pure unstable particle $\alpha$, moving observers may see also its decay products $\beta + \gamma$ with some probability.  We can prove an even stronger statement: If all (both resting and moving with different rapidities $\boldsymbol \theta$) observers see the unstable particle $\alpha$ at $t=0$ with 100\% probability

\begin{align}
\Upsilon(\theta, 0) =1 \label{eq:omega-theta-1}
\end{align}

\noindent then this particle is stable with respect to time translations as well.

Suppose that Eq. (\ref{eq:omega-theta-1}) is true, i.e., for any $ | \Psi \rangle \in \mathscr {H} _ {\alpha} $

\begin{align*}
\Upsilon(\theta,0) &= \left\langle \Psi \left| e^{-\frac{ic}{\hbar}K_x \theta} T e^{\frac{ic}{\hbar}K_x \theta} \right| \Psi \right\rangle = 1
\end{align*}

\noindent This means that all boosts leave the subspace $ \mathscr {H} _{\alpha} $  invariant

\begin{align*}
e^{\frac{ic}{\hbar}K_x \theta} | \Psi \rangle \in \mathscr{H}_{\alpha}
\end{align*}

\noindent  and that the interacting boost operator $ K_x $ commutes with the projection $ T $. Then  commutators (\ref{eq:5.55}), (\ref{eq:commp0}) and the Jacobi identity imply

\begin{align*}
[T,H] &= -\frac{ic^2}{\hbar}[T,[K_x,P_{0x}]] \\
&= \frac{ic^2}{\hbar}
[K_x,[P_{0x},T]] + \frac{ic^2}{\hbar}[P_{0x},[T,K_x]] = 0
\end{align*}

\noindent which contradicts the fundamental property  (\ref{eq:comm}) of unstable particles. To resolve this contradiction, we have to admit that the boosted state  $e^{\frac{ic}{\hbar}K_x \theta} | \Psi \rangle$ does not correspond to the particle $\alpha$ with 100\% probability. This state must contain an admixture of decay products even at the initial time $t=0$

\begin{align}
e^{\frac{ic}{\hbar}K_x \theta} | \Psi \rangle &\notin \mathscr{H}_{\alpha}
 \nonumber \\
\Upsilon(\theta, 0) &< 1, \mbox{  } for \; \; \theta \neq 0
\label{eq:not-in2}
\end{align}

\noindent This is the ``decay caused by boost'' \cite{Stefanovich_decay, Stefanovich_dec, Giacosa}, which means, among other things, that special-relativistic formulas  (\ref{eq:ups-theta}) and (\ref{eq:omega-theta-1}) are inaccurate, and that boosts have a nontrivial effect on the internal state of the unstable particle.

\subsection{Discussion}

Here we discussed the dynamical effect of boosts on unstable particles (\ref{eq:not-in2}). However, similar non-traditional effects should be visible also in other interacting systems, even in classical (non-quantum) ones \cite{Grobe2}.
In order to verify these predictions, one has to look at composite interacting systems, where interaction acts for a sufficiently long time. Unfortunately, most experimental checks of
special relativity \cite{experimentSR, MacArthur, Roberts} do not satisfy these criteria.  For example, dynamical boosts do not change the relativistic kinematics (the relationships between momenta, velocities and energies of free particles) in collisions, reactions and decays.
Likewise, dynamical boosts do not affect Doppler type experiments \cite{Kaivola, Hasselkamp}, which measure the frequency (energy) of light and its dependence on the motion of the source or the observer.
Michelson-Morley type  experiments \cite{Muller, Wolf, Alvager}, studying the invariance of the speed of light, are not affected as well.

The time dilation experiments with unstable  particles \cite{dilation, Frisch, dilation1, dilation2} are exceptional, because they study systems that are under the action of interaction during sufficiently long time. Unfortunately, predicted deviations from the special-relativistic time dilation formula (\ref{eq:ups-theta}) are too small to be observed.  One can also see that the ``decay caused by boost'' effect is also very small and beyond the capabilities of modern techniques \cite{Stefanovich_dec}.

Perhaps, the most convincing evidence for the dynamical character of boosts was obtained in the Frascati experiment \cite{Calcaterra, Pizzella2, Pizzella-2017}, which established the superluminal dynamics of the electric field of relativistic charges.  This observation was explained from the point of view of the Poincar\'e-Wigner-Dirac theory in  \cite{mybook, volume2, Stefanovich-2017}.

\section{Conclusions}

We applied Poincar\'e-Wigner-Dirac theory of relativistic interactions to unstable particles.  In particular, we were interested in how the same particle is seen by different moving observers. We proved that the decay probability cannot be invariant with respect to boosts. Different moving observers may see different internal compositions of the same particle.  Unfortunately, this effect is too small to be visible in modern experiments.


%

\end{document}